# Usage of Cloud Computing Simulators and Future Systems In For Computational Research

Dr. Ramkumar Lakshminarayanan*[1]
Mr. Rajasekar Ramalingam *[2]

[1][ramkumar.sur@cas.edu.om] [2][rajasekar.sur@cas.edu.om]
*Faculty, Department of Information technology, College of Applied Sciences – Sur, Oman.

## ABSTRACT

Cloud Computing is an Internet based computing, whereby shared resources, software and information, are provided to computers and devices on-demand, like the electricity grid. Currently, IaaS (Infrastructure as a Service), PaaS (Platform as a Service) and SaaS (Software as a Service) are used as a business model for Cloud Computing. Nowadays, the adoption and deployment of Cloud Computing is increasing in various domains, forcing researchers to conduct research in the area of Cloud Computing globally. Setting up the research environment is critical for the researchers in the developing countries to evaluate the research outputs. Currently, modeling, simulation technology and access of resources from various university data center's has become a useful and powerful tool in cloud computing research. Several cloud simulators have been specifically developed by various universities to carry out Cloud Computing research, including CloudSim, SPECI, Green Cloud and Future Systems (the Indiana University machines India, Bravo, Delta, Echo and Foxtrot) supports leading edge data science research and a broad range of computing-enabled education as well as integration of ideas from cloud and HPC systems. In this paper, the features, suitability, adaptability and the learning curve of the existing Cloud Computing simulators and Future Systems are reviewed and analyzed.

Keywords: Cloud Computing, Simulators, Future System

## I. INTRODUCTION

Cloud Computing is a prominent area of research, as there is a challenge and need to increase the productivity and performance in the organization with the increasing demand of infrastructure, software and platform [1]. Initially, research labs, companies and universities used data centers, a large cluster of networked computer servers typically utilized in the processing, distribution of data and remote storage [2].

The drawbacks in the data center are (1) no possibility for infinite computing resources on demand, (2) no upfront commitment, (3) no pay for use of computing resources, (4) no scaling, (5) no possibility of multiplexing of workloads and (5) no resource virtualization [3]. Researchers have initially coined the term Grid Computing to avail the computing power on demand similar to the usage of electric power grid. Earth System Grid, Open Science Grid, caBig, TeraGrid and Earth System Grid are notable research projects [4]. Commercial ventures are not emerged on the Grid Computing, but emerged in Cloud Computing. In Cloud Computing, Infrastructure (IaaS), Platform (PaaS) and Software (SaaS) are delivered as a subscription based pay-as-you-go model to consumers [5]. The key difference between the Cloud Computing and the Grid Computing is virtualization and service level agreement. Infrastructure as a service (IaaS) provides virtual infrastructure, virtual storage and computer hardware, e.g. Amazon AWS [6], and Rackspace Open Cloud [7]. Platform as a Service (PaaS) provides the developing platform supporting the full Software Development Life Cycle (SDLC) e.g. Windows Azure [8] and Google App Engine [9]. Software as a Service (SaaS) provides the user interface and application management, e.g. Google Docs [10] and Microsoft Office 365 [11].

The characteristics of Cloud Computing [12] [13] are:
- On-demand self-service: Without the human association with cloud administration the resources are gained and utilized on the requirement
- A broad network access: Availing the resources in the wider range of devices.
- Resource pooling: Pooling of resources to numerous client.
- Rapid elasticity: Scaling out and can scale back.
- Measured service: Cloud Resources utilization is measured by monitoring storage usage, CPU hours, and bandwidth usage

The research in the area of Cloud Computing is i.e. 1879 publications during the block year 2009-2013. In 2013, there are 803 publications compared to 55 in 2009 witnessing a tremendous growth [14].

The major categories of research issues in Cloud Computing are Cloud Performance (CP), Data Management (DM), Data center management (DCM), Software Development (SD) and Service Management (SM) [15].

1. *Cloud Performance* focus on the evaluation and optimization of the performance of the clouds. Evaluation and optimization of the performance of the clouds include studies on (1) quantifying and comparing performance among different clouds [16], (2) efficient workflow scheduling and load balancing [17], (3) optimizing resource allocation [18], (4) identifying anomalous and enabling automatic bottleneck detection [19], (5) prediction of node failure [20], and (6) improving interoperability across different clouds [21].

2. *Data Management* focus on issues associated with the large scale, distributed data processing in the clouds. It includes (1) demand tradeoff between availability and data consistency [22], (2) minimizing the data redundancy while achieving a given high data reliability [23], (3) integrating and analyzing data from geographically distributed data sources [24], (4) reducing query execution time [25], and (5) improving the performance of transaction processing [26].

3. *Data Center Management* researches focus on energy efficiency, power conservation, and environmental considerations in the design and architecture of data centers [27]. In addition, proposed energy-aware scheduling algorithms [28].

4. *Software Development* category represents a stream of software developer-oriented research. It includes anticipating runtime problems in the cloud and defining metrics for better fault localization [29], identifying issues related to scalability, availability, data integrity, data transformation, data quality, data heterogeneity, privacy, legal and regulatory issues, and governance with respect to Hadoop [30], component-based approach to computational science applications on cloud infrastructures [31] and automating the application distribution to multiple hosts [32].

5. *Service Management* focuses on the administration of Cloud Computing services. The researches includes challenges and issues in the service lifecycle of the cloud [33] and selecting, discovering, and publishing in cloud-based services [34].

The major challenge for the researcher are the need for the environment to conduct the researches in the above discussed categories as the cost involved in the conducting the research in the real cloud is relatively high. It is really important to identify the suitability and applicability of the simulator and university data centers.

## II. BACKGROUND

This study investigates various research works in the area of Cloud Computing to identify the cloud simulator usage in the fields. There are similar works in this area to survey the future direction of the cloud simulators [35] and comparative study in the Cloud Computing [36]. As most of the simulators and universities data center are open source and developed to support their research projects, the simulator may not be up-to-date and university data center would have stopped support their project for research. In our paper, we have discussed the features of the commonly used simulator for research. At the end, we have provided a summary to compare the simulators and university data centers for their capabilities and suitability.

## III. CLOUD SIMULATORS

A. CLOUDSIM

CloudSim is a toolkit (library) developed by CLOUDS Laboratory at the Computer Science and Engineering Department of the University of Melbourne, Australia for Cloud Computing Simulation.
The feature in simulator provides the required classes to describe data centers, virtual machines, applications, users, computational resources, and policies. The multi layered design of the Cloud Computing architecture is shown in Figure 1 [37]. In the system level, the collection of data centers generally hundreds to thousands of hosts provides the computing power in cloud environments.  Core Middleware (PaaS) layer provides run-time environment by implementing the platform-level services and as well as accessed by the SaaS and IaaS layers. User Level Middleware layer has the software framework that helps the developers in creating user-interfaces, deployment and execution of applications in the clouds.  There is a support for multi-layer applications development, such as Hibernate and spring.

The researches supported by CloudSim are:

- Modeling and simulation of large scale Cloud Computing data centers
- Modeling and simulation of virtualized server hosts, with customizable policies for provisioning host resources to virtual machines
- Modeling and simulation of energy-aware computational resources

- Modeling and simulation of data center network topologies and message-passing applications
- Modeling and simulation of federated clouds
- Dynamic insertion of simulation elements, halt and resume of simulation
- User-defined policies for allocation of hosts to virtual machines and policies for allocation of host resources to virtual machines

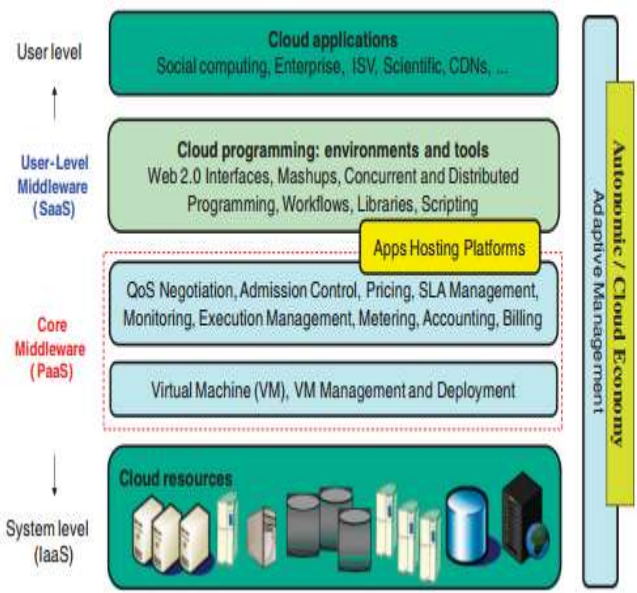

Figure 1. Layered Cloud Computing Architecture.

In this toolkit by extending or replacing the classes and codes, a new scenario can be designed for research. By the setting the various parameters, the desired environment can be set for analysis.

Based on the requirement, CloudSim supports the researchers in developing additional extensions added to its existing functionality. Table 1 shows the list of extensions and their functionality. The Cloudsim projects and its extensions are up-to-date and Active.

**Table 1. Cloudsim extensions and their functionality**

| S.no | Extension | Features |
|---|---|---|
| 1. | CloudSimEx | Web session modelling, logging utilities, generating CSV file, multiple experiments in parallel and MapReduce simulation. |
| 2. | CloudSim Automation | Automate cloud simulation, reuse, extension and sharing of simulation scenarios. |
| 3. | WorkflowSim | Workflow parser, workflow engine and job scheduler. |
| 4. | Cloud2Sim | Distributed concurrent architecture to CloudSim simulations |
| 5. | Simple Workflow | Simulating workflows on CloudSim |
| 6. | DynamicCloudSim | Includes Models for Heterogeneity in performance, uncertainty in the performance of virtual machines. |
| 7. | RealCloudSim | Allocation of virtual machines based on main engine |
| 8. | CloudReports | Report generation features and creation of extensions in a plugin fashion |
| 9. | CloudAuction | Auction-based mechanisms in CloudSim |
| 10. | CloudMIG Xpress | The migration of software systems to PaaS or IaaS-based Cloud environments |
| 11 | Cloud Analyst | Support evaluation of social network tools according to geographic distribution of users and data centers. |

B. SPECI

SPECI, a simulation tool developed for the exploration of scaling properties of large data centers. Given the middleware design policy as input, it simulates the performance and behavior of the data center. Using this simulator, when the number of nodes involved scales up, it is possible to observe the overall system behavior under this protocol. The most common result obtained from this tool are:

- With the given settings and size, the overall network load of the protocol is evaluated.
- Time-for-consistency for the system can be retrieved [38].

The SPECI simulator is not up-to-date and not active.

C. GREEN CLOUD

Cloud computing operates relying on a pool of shared computing resources available on demand and usually hosted in data centers. Assessing performance and energy efficiency of data centers become fundamental. Industries use a number of metrics to assess efficiency and energy consumption of Cloud Computing systems, focusing mainly on the efficiency of IT equipment's, cooling and power distribution systems.

The categories of existing metrics to access, efficiency, performance and quality of Cloud Computing systems are:
- Energy efficiency and Power
- Environment and air management
- Cooling efficiency

Green Cloud supports the research in the above categories [39]. Green Cloud Simulators are up-to-date and Active.

D. FUTURE SYSTEMS

Future Systems. the Indiana University machines India, Bravo, Delta, Echo and Foxtrot support leading edge data science research and a broad range of computing-enabled education as well as integration of ideas from cloud and HPC systems. The distributed testbed Future Systems allow users to perform experiments with cloud, grid and high performance. To determine the experiments impact on the infrastructure, future systems gather performance information.

The intent of Future System are:
- To provide a heterogeneous hardware environment
- To deploy heterogeneous hardware
- To deploy configurable environment
- Tools for experiments on the infrastructure.

The Future Systems heterogeneous cluster is connected by high performance network capable of providing the environment and IaaS Cloud. Future Systems provide a component Cloud mesh [40] to deliver a software-defined system. Cloud mesh encompasses virtualized and bare-metal infrastructure, networks, application, systems and platform software. It provides cloud Testbeds as a Service (CTaaS). Future Systems Cloud Mesh is active and up-to-date and supports for researches by approval.

**Table 2 Lists the comparison of cloud usage.**

| S. No | Name | Base Platform | Energy Model | Area of Research |
|---|---|---|---|---|
| 1 | CloudSim | SimJava | Yes | CP, DM, DCM, SD and SM |
| 2 | SPECI | SimKit | Partial | SD and SM |
| 3 | Green Cloud | NS2 | Yes | DCM |
| 4 | Future Systems (CloudMesh) | - | Yes | CP, DM, DCM, SD and SM |

**Table 2.Comparison of Cloud Usage**

CloudSim supports the researches in the area of Cloud Performance, Data Management, Data center Management, Software Development and Service Management. SPECI supports the researches in area of Software Development and Service Management where as it is not suitable for the researches in the area of Cloud Performance, Data Management and Data center Management. Green Cloud is suitable for the Data center Management researches. Future Systems is suitable for all research areas.

IV. Conclusion

This paper compared the emerging simulators and cloud platform that pursued the advance scientific research of cloud computing. Among the cloud simulators CloudSim is up-to-date and it is an active project which supports research in major categories. SPECI is suitable for researches of software development and service management categories, but the SPECI related researches are not up-to-date. The Future Systems Cloud Mesh also suitable for all the research categories. Green Cloud is suitable for researches focusing energy efficiency and power conservation.